\documentclass[twocolumn,showpacs,pre,aps,superscriptaddress]{revtex4}
\usepackage[utf8]{inputenc}
\usepackage[english]{babel} 
\usepackage{graphicx} 
\usepackage{listings} 
\usepackage{amssymb} 
\usepackage{amsmath} 
\usepackage{longtable}
\usepackage{multirow} 
\usepackage{booktabs}
\usepackage{pst-all} 
\usepackage{pstricks-add}
\usepackage{units} 
\usepackage{listings} 
\usepackage{color}
\usepackage{subfigure}

\begin{document}

\title{Macroscopically deterministic, Markovian thermalization in finite quantum spin systems}

\author{Hendrik Niemeyer}
\email{hniemeye@uos.de}

\affiliation{Fachbereich Physik, Universit\"at Osnabr\"uck,
             Barbarastrasse 7, D-49069 Osnabr\"uck, Germany}

\author{Kristel Michielsen}
\email{k.michielsen@fz-juelich.de}

\affiliation{Institute for Advanced Simulation, J\"ulich Supercomputing Centre,\\
Forschungszentrum J\"ulich, D-52425 J\"ulich, Germany and
RWTH Aachen University, D-52056 Aachen,
Germany}
             
\author{Hans de Raedt}
\email{h.a.de.raedt@rug.nl}

\affiliation{Department of Applied Physics, Zernike Institute for Advanced Materials,\\
University of Groningen, Nijenborgh 4, NL-9747AG Groningen, The Netherlands}

\author{Jochen Gemmer}

\email{jgemmer@uos.de}

\affiliation{Fachbereich Physik, Universit\"at Osnabr\"uck,
             Barbarastrasse 7, D-49069 Osnabr\"uck, Germany}
             
\date{\today}

\begin{abstract}
A key feature of non-equilibrium thermodynamics is the Markovian, deterministic relaxation of coarse observables such as, for example, the
temperature difference between two macroscopic objects which evolves independently of almost all details of the initial state.
We demonstrate that the unitary dynamics for moderately sized spin-1/2 systems may yield the same type of relaxation dynamics for a
given magnetization difference. This observation might contribute to the understanding of the emergence of thermodynamics within
closed quantum systems.
\end{abstract}


\maketitle
\section{introduction}

Roughly 100 years after its systematic microscopic interpretation the origin of thermodynamics is still under dispute (see e.g.,
\cite{Butterfield2007, Popescu2006, Beretta2010} and references therein). It is, however,
an empirical fact
that macroscopic systems behave according to the laws of thermodynamics and are routinely viewed as large quantum systems.
Consequently, already in the early formulations of
quantum mechanics \cite{Pauli1928, Goldstein2010, Schroedinger1948, VanHove1955, Tasaki1998} the question about the relationship
between quantum mechanics and thermodynamics arose and is still discussed today.
A central point in the discussion is the reconciliation of unitary quantum dynamics (featuring no fixed point) with the equilibrating, rate equation-type
dynamics of non-equilibrium thermodynamics (featuring  a fixed point).
In this debate various concepts have been introduced  such as
``typicality'' \cite{Goldstein2010, Page1993, Goldstein2006, Popescu2006,  Gemmer2003, Reimann2007},
``pure state quantum statistical mechanics'' \cite{Reimann2008, Linden2009, Riera2012}
``eigenstate thermalization hypothesis'' \cite{Goldstein2010, Deutsch1991, Srednicki1994, Rigol2008, Dubey2012}, ``thermal environment
coupling'' \cite{Alicki1987,Caldeira1983,JIN10X}, and many more.
Recently, experiments in an optical lattice with ultra-cold atoms
have been performed to study the relaxation dynamics in an interacting many-body system \cite{Trotzky2012}.
Also the thermalization dynamics itself, beyond the mere existence of equilibrium, has gained attention:
Fokker-Planck equations for some closed finite quantum systems have been suggested \cite{Tikhonenkov2013,Lesanovsky2012,Ji2013, Niemeyer2013}.

In this paper we discuss a quantum model corresponding to the archetypical thermodynamic scenario in which
two (equal or similar) macroscopic bodies
are prepared at, e.g., different temperatures (possibly a hot and a cold coffee mug) and then brought into contact
but kept isolated from any environment. (Similar scenarios have been analyzed using quantum models in, e.g., Refs. 
\cite{Ponomarev2011, Zhang2012}.) Experimental evidence shows that the dynamics of the temperature
difference (here called $x$) is autonomous and
Markovian in the sense that it may be described as
\begin{equation}
  \dot{x}  = -R(x)  x,
\label{mactemp}
\end{equation}
where $R$ denotes the rate of change.
This implies that the dynamics of the temperature difference $x$ possesses a unique, attractive fixed point, is
free of memory effects and is not affected by any other variables.
If the temperature difference is in some sense a statistical quantity
then its variance $\sigma^2=\langle x^2 \rangle - \langle x \rangle^2$
is expected to be small compared to the overall scale of its expectation value $\langle x \rangle$, a property referred to as
macroscopic determinism.
In terms of the coffee mugs this means that one may repeat the experiment several times without getting measurably different results for
$x$ during its evolution.

\section{model and observables}

The quantum model we consider is a finite, anisotropic Heisenberg spin-ladder of size $N$ described by the
Hamiltonian ($\hbar=1$ throughout this paper)
\begin{align}
\hat{H} &= \hat{H}_0 +\kappa \hat{V} \nonumber \\
  \hat{H}_0  &= \sum_{i=1}^{N/2-1}\sum_{\alpha=L,R} J(\hat{S}_x^{\alpha,i} \hat{S}_x^{\alpha, i+1} + \hat{S}_y^{\alpha,i} \hat{S}_y^{\alpha, i+1} + 0.6 \; \hat{S}_z^{\alpha,i}
\hat{S}_z^{\alpha, i+1}) \nonumber \\
   \hat{V} & =\sum_{i=1}^N (\hat{S}_x^{L,i} \hat{S}_x^{R,
i} + \hat{S}_y^{L,i} \hat{S}_y^{R, i} + 0.6 \;
\hat{S}_z^{L,i} \hat{S}_z^{R, i}),
\label{model}
\end{align}
where $\hat{S}_{x}$, $\hat{S}_{y}$ and $\hat{S}_{z}$  denote the spin-1/2 operators, $J=1$ is the coupling strength along the beams of
the ladder (labeled by $\alpha=L,R$) and $\kappa=0.2$ is the coupling strength along the rungs.
The observable which is our analogon to the temperature
difference mentioned in the example of the two coffee mugs is the difference of magnetization along the z-axis between the
two beams which we call $\hat{x}$ throughout the paper
\begin{equation}
 \hat{x}=\sum_i \hat{S}_z^{L,i} - \hat{S}_z^{R,i}
 \label{defmagdiff}.
\end{equation}
Note that in the model described by Eq.~(\ref{model}) the total magnetization along the z-axis $\hat{S}_z^{total}$ of the entire system
is a conserved quantity. Hence, for our analysis we choose the largest total magnetization subspace $S_z^{total}=0$. Within this subspace
the eigenvalues of $\hat{x}$ are $X=-N/2, -N/2+2,\cdots, N/2-2, N/2$. The multiplicities are essentially binomially distributed, $X=0$ 
features the largest degeneracy.

In Ref. \cite{Niemeyer2013} this model has been  analyzed for
$N=16$ by means of exact diagonalization. Reasonable agreement of the quantum dynamics of $x$ with a Fokker-Planck equation was found
numerically for a small set of initial states all of which are in a sense close to equilibrium.
Furthermore the respective Fokker-Planck equation has been ``derived'' from an appropriate projection operator technique (up to leading order) 
under the assumption of equal correlation times for the transition dynamics between all $X$-subspaces. In order to investigate the claim that
this finite quantum model would yield irreversible Markovian $x$-dynamics for all practical purposes, we numerically analyze the same model 
class in the paper 
at hand but for larger systems and a much wider range of initial states. We essentially find that, while indeed autonomous Markovian, 
deterministic $x$-dynamics emerges in general, specific predictions of the Fokker-Planck model suggested in 
\cite{Niemeyer2013} fail for initial states further away from equilibrium. Rather than to vanish, this failure appears to become 
even more pronounced for larger systems.  Thus, below we present an alternative analysis based on typicality rather than on
projection operator techniques which at least predicts specific equilibrium values correctly while being methodically sound.

In this paper we focus on spin systems. Relaxation in closed quantum systems is, however, not limited to spin systems, for an example 
of a bosonic system, see, e.g., Ref. \cite{Zhang2011}.

\section{computational scheme and initial states }

As mentioned above, in the present paper we address larger systems and a larger variety of different pure, rather than mixed,
initial states.
To those ends we solve the time-dependent Schr\"odinger equation (TDSE) numerically by means
of the Chebyshev polynomial algorithm~\cite{TALE84,LEFO91,IITA97,DOBR03}.
This algorithm yields results that are very accurate
(close to machine precision), independent of the time step used~\cite{RAED06}.
Conserved quantities such as the total energy and magnetization
are constant to almost machine precision (about 14 digits in our calculations).
Computer memory severely limits the sizes of the quantum spin systems
which can be simulated.
To represent the state $\left|\Psi(t)\right\rangle$ of $N$ spin-$1/2$ particles on a digital computer,
we need a least $2^{N+4}$ bytes.
In practice, we need several of such vectors, memory for communication buffers,
local variables and the code itself.
For example, for $N=32$ we need about 320 GB of memory.
Although the CPU time required to solve the TDSE also increases exponentially with the number of spins,
this increase can be compensated for by distributing the calculations over many processors.
For a $N=32$ system, solving the TDSE upto $t=150$
using 65536 CPUs takes about 6 hours on the J\"ulich IBM BlueGene/Q.
Details of the massively parallel simulation code are given in Ref.~\cite{RAED07X}.

To account for the initial state independence as described in the introduction, we draw pure states essentially at random,
only tailored to feature probability distributions with respect to $x$ and energy that are narrow compared to the overall
range of possible values for
the respective observables. From the numerics it turns out that the narrow energy distribution is crucial for the expected
dynamics to emerge. If the initial states are drawn from a larger energy window the equilibrium  variances $\sigma^2 $ exhibit a strong 
dependence  on the initial state which is obviously in conflict with the concept of a thermodynamic equilibrium. Our understanding of this 
phenomenon is not conclusive yet, we however expect that its occurrence strongly depends on the degree to which the eigenstate 
thermalization hypothesis is fulfilled. We intend to discuss this thoroughly in a forthcoming paper \cite{Steinigeweg2013}.  
The initial states $|\omega_X\rangle$ are constructed as follows. We begin with the states
$|\Psi\rangle=\sum_{j=1}^{2^N} c_j\left|j \right\rangle$,
%
where the set of states $\left\{\left|j\right\rangle\right\}$ denotes
the complete basis set of states in the spin-up -- spin-down representation
and the coefficients $c_j$ are obtained by generating uniform independent random numbers in the interval $[-1,1]$
and rescaling them such that $\sum_{j=1}^{2^N} |c_j|^2 = 1$. We then project the initial state
to the ``$S_z^{total}=0$-and-specific-$X$-subspace'' and in order to narrow down the energy distribution
(in this case around $\langle \hat{H} \rangle = 0$) we eventually apply a pertinent exponential
\begin{equation}
 |\omega_X\rangle =  C e^{-\alpha \hat{H}^2} \hat{P}_{x}  \hat{P}(S_z=0) |\Psi\rangle,
\label{inist}
\end{equation}
with $\hat{P}_{x}$ being the projector onto a subspace featuring a certain
eigenvalue of $\hat{x}$,
$\hat{P}(S_z=0)$ being the projector on the $S_z^{total}=0$ subspace and $\alpha$
denoting a constant chosen such that the variance of the energy $\sigma_H^2$ for the initial states $\omega$ is small ($\sigma_H = 0.37$
is used throughout this paper for the reasons explained above). $C$ is just a normalization constant.
To compute $e^{-\alpha \hat{H}^2}|\phi\rangle$ the same Chebyshev polynomial algorithm is used as for the
time evolution. Of course, using this algorithm the energy window can also be located at positions other than $\langle \hat{H} \rangle = 0$. 
In this paper, however, we choose  $\langle \hat{H} \rangle = 0$ since it appears to be the most promising choice for the emergence of 
thermodynamical behavior: In this model $\langle \hat{H} \rangle = 0$ is in the center of the full energy spectrum and features the largest
density of states. 
Furthermore 
 $\langle \hat{H} \rangle = 0$ would also be the energy expectation value of a hypothetical Gibbs equilibrium state with infinite 
 temperature. So in this respect the choice $\langle \hat{H} \rangle = 0$ corresponds to a high temperature limit. The case 
 of  $\langle \hat{H} \rangle$ closer to the ground state energy, i.e., lower temperatures is left as an interesting subject for 
 further investigation. 
 
\section{discussion of the dynamics}

As a first result we find that initial states with the same $\langle\hat{x}(0)\rangle$ and $\alpha$ but generated
from  different random $|\Psi\rangle$ show approximately the same dynamics of, e.g., mean $\langle \hat{x}(t)\rangle$ and variance 
$\sigma^2(t) = \langle \hat{x}(t)^2 \rangle  - \langle \hat{x}(t) \rangle^2$. 
This was also found and discussed in Ref.~\cite{Niemeyer2013} for $N=16$.
However, for $N=32$ this becomes so pronounced that the respective graphs cannot be discriminated by the naked eye.
Therefore, in what follows we only present results for initial states generated from one random $|\Psi\rangle$.
This finding may be viewed as a manifestation of the concept of dynamical
typicality \cite{Bartsch2009} and is in accord with results presented in Refs.~\cite{HAMS00,Elsayed2013}.

\begin{figure}[t]
\centering
 \includegraphics[width=7cm, height=5cm]{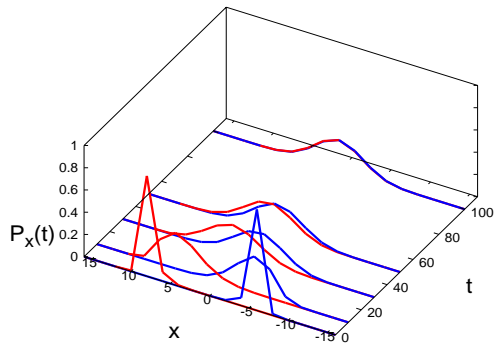}
 \caption{(color online). Dynamics of the probability distribution  ${P}_x(t)$ for spin-ladders of size $N=32$ for two different essentially 
 random
initial states with $\langle \hat{x}(0) \rangle=8$ (red line) and $\langle \hat{x}(0) \rangle=-6$ (blue line).
The two probability distributions hardly overlap at $t=0$ and almost coincide at later times.}
 \label{fig:fancy}
\end{figure}

Next we turn towards an overall
picture of the $x$-dynamics in this model. Fig.~\ref{fig:fancy} shows
the dynamics of the probabilities $P_x=\langle \hat{P}_x\rangle$ for two
initial states with $\langle \hat{x}(0) \rangle=8$ and $\langle \hat{x}(0) \rangle=-6$, respectively, for  $N=32$. 
Obviously the unitary dynamics yields after some time essentially constant probabilities $P_x(t)$ which coincide for both initial states
while they hardly overlap at $t=0$. This can be interpreted as a strong indication for thermodynamic behaviour in this spin system.
To analyze
this further we focus on the dynamics of the magnetization difference $\langle \hat{x}(t) \rangle$ and its variance
$\sigma^2(t) = \langle \hat{x}(t)^2 \rangle  - \langle \hat{x}(t) \rangle^2$. 

In Ref.~\cite{Niemeyer2013} it has been argued that the $x$ dynamics of the model could be captured by a Markovian master equation derived 
from a simple stochastic spin-flip model. The transition rates between neighboring $x$-subspaces read 
\begin{equation}
R_{(X \rightarrow X \pm 2)}=\frac{\gamma \kappa^2 N}{2}\left(\frac{1}{2} \mp \frac{X}{N}\right)^2, 
\label{rates}
\end{equation}
where $\gamma$ denotes an overall time constant.

\begin{figure}[t]
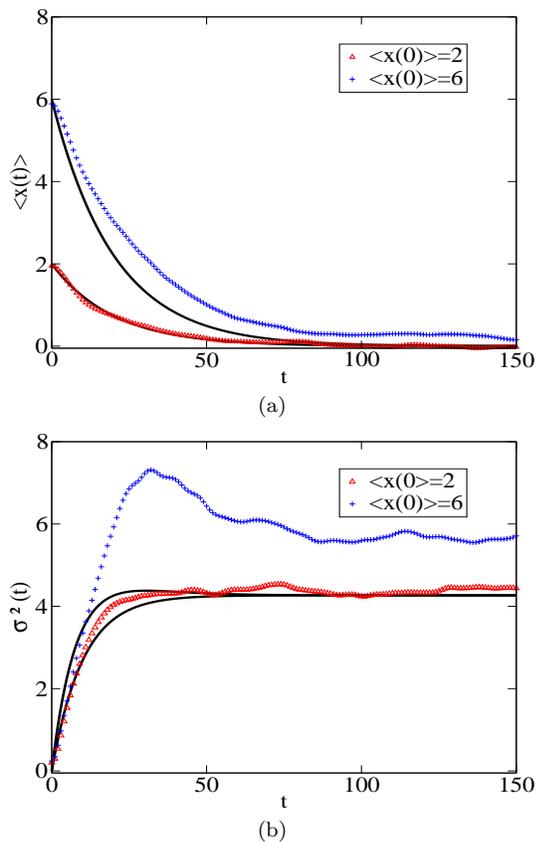

\centering
\subfigure[]{
   \includegraphics[width=7cm,height=5cm]{fig2a.eps}
\label{fig:rate16}}
\subfigure[]{
\includegraphics[width=7cm,height=5cm]{fig2b.eps}
\label{fig:rate32}}
\caption{(color online). (a) Expectation values $\langle\hat{x}(t) \rangle$ of the magnetization difference 
for $N=16$ for two states, one starting close to equilibrium $\langle \hat{x}(0) \rangle =2$ and the other far from equilibrium  
$\langle \hat{x}(0) \rangle = 6$. Solid lines represent the corresponding data as calculated from the stochastic model suggested in 
Ref.~\cite{Niemeyer2013}. Close to equilibrium the agreement is good while more off-equilibrium it is not. (b) Equivalent data and 
style of presentation as in (a) but for the variances $\sigma^2(t)$.}
\label{fig:N16}
\end{figure}

\begin{figure}[t]
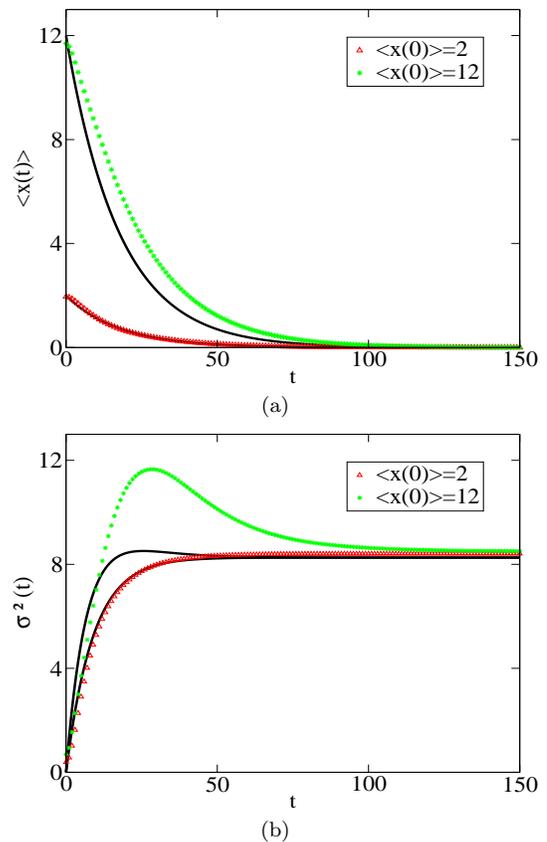

 \centering
\subfigure[]{
   \includegraphics[width=7cm,height=5cm]{fig3a.eps}
\label{fig:rate1632}}
\subfigure[]{
\includegraphics[width=7cm,height=5cm]{fig3b.eps}
\label{fig:rate3232}}
\caption{(color online). (a, b) Equivalent data and 
style of presentation as in Fig. \ref{fig:N16} but
for $N=32$. ( $\langle \hat{x}(0) \rangle = 12$ here represents off-equilibrium ). While the agreement close to equilibrium is better 
than for $N=16$, for off-equilibrium initial states this appears not to be the case.}
\label{fig:N32}
\end{figure}

Master equations with rates (\ref{rates}) necessarily yield autonomous dynamics. The $x$-dynamics as resulting from the  Schr\"odinger 
equation and 
from the above master equation are compared to each other for 
two different initial states for $N=16$, see Fig.~\ref{fig:N16}. While the agreement for the initial state close to equilibrium is quite good 
($\langle \hat{x}(0) \rangle=2$) the stochastic model (\ref{rates}) fails to predict the dynamics for states starting far from equilibrium 
($\langle \hat{x}(0) \rangle=6$). In Ref.~\cite{Niemeyer2013} it was suggested
that this 
failure may vanish for larger system sizes. With the work at hand we are able to address such larger system sizes. Data 
equivalent to  Fig.~\ref{fig:N16} but now for $N=32$ is given in Fig.~\ref{fig:N32}. Obviously the agreement for initial states close to 
equilibrium ($\langle \hat{x}(0) \rangle=2$) becomes even better but the failure for substantially off-equilibrium initial states   
($\langle \hat{x}(0) \rangle=12$) remains. Since the agreement of  (\ref{rates}) with the true quantum dynamics does not improve for
off-equilibrium states with increasing system size the crucial question remains whether the true quantum dynamics may nevertheless be considered in 
accord with macroscopically deterministic, irreversible, Markovian behavior of $x(t)$ for larger $N$. 

\begin{figure}[t]
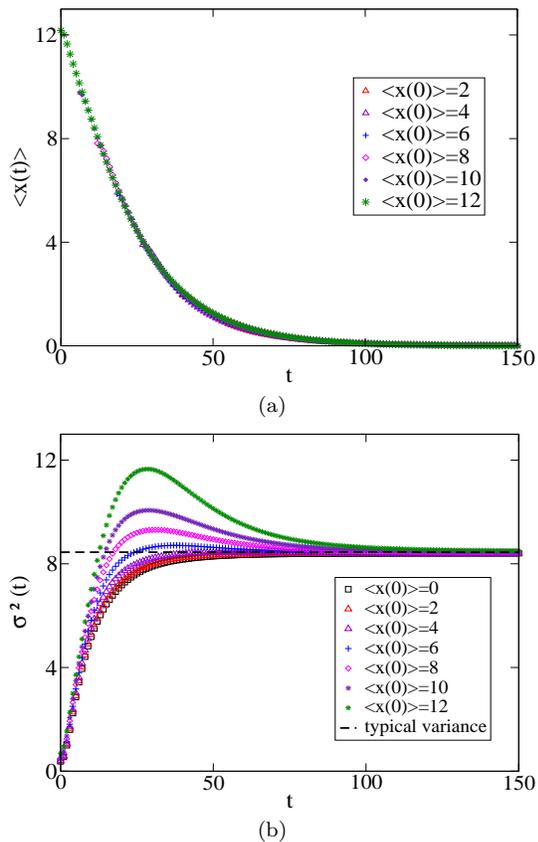

 \centering
\subfigure[]{
   \includegraphics[width=7cm,height=5cm]{fig4a.eps}
\label{fig:mean32}}
\subfigure[]{
\includegraphics[width=7cm,height=5cm]{fig4b.eps}
\label{fig:var32}}
\caption{(color online). (a) Expectation values $\langle\hat{x}(t) \rangle$ of the magnetization difference for $N=32$ and 
initial states featuring different $\langle \hat{x}(0) \rangle$. Graphs are shifted in time for optimal agreement. (b) Variances
$\sigma^2$ of the magnetization difference for $N=32$ and initial states featuring different
$\langle \hat{x}(0) \rangle$ (unlike (a), no time shift is applied).
The dashed line indicates the typical variance (see text).}
\label{fig:n32}
\end{figure}

In order to address this question we compute $\langle \hat{x}(t) \rangle$ for various  $\langle \hat{x}(0) \rangle$ for $N=32$ and 
display the 
result in  Fig.~\ref{fig:mean32}. The curves are shifted in time such that the squares of the deviations of the curves from one another
are minimized. If the dynamics was fully autonomous and Markovian, the curves
would lie exactly on top of each other. Apparently, this is to good accuracy the case for all initial states, regardless of the failure of  
(\ref{rates}). In  Fig.~\ref{fig:var32} the 
variances $\sigma^2(t)$  are displayed for the same initial states. Although the master equation based on (\ref{rates}) does not describe the 
dynamics correctly, all
variances appear to converge to the same value. This finding is also in accord with Markovian irreversible behavior. 

How can this ``irreversible'' tendency of the variances corresponding to so many pure, different initial states towards one 
constant ``equilibrium'' value be explained although the description in terms of the master equation defined by (\ref{rates}) fails?
An explanation may be provided by the concept of typicality. The equilibrium value coincides with the typical variance
$\sigma^2_{typical}$, see Fig.~\ref{fig:var32}. The latter is the ``generic'' variance of
a random initial state which is unrestricted with respect to $\langle \hat{x}\rangle$, i.e. 
$|\omega'\rangle = e^{-\alpha \hat{H}^2} |\Psi\rangle$ (cf. also \cite{Sugiura2012}):
Within the addressed energy shell there are certainly states featuring variances ranging from $\approx 0.5$ to $\approx N^2/16$.
However, as shown in the context of typicality, states featuring a certain variance $\sigma^2_{typical}$
are by far the most frequent ones, with respect to the unitary invariant measure \cite{Bartsch2009}. Therefore they are sometimes called
``typical states''. While all considered initial states start in a very tiny region of
Hilbert space formed by ``non-typical'' states, some states venture out into
the extremely large region formed by the typical states, while other initial states do not such as, e.g., the initial state coresponding to 
 $\langle \hat{x}(0) \rangle = 6 $ in  Fig. \ref{fig:rate32}. Its variance  $\sigma^2(t)$ remains significantly above the 
typical variance. 
However, the corresponding data for $N=32$ as displayed in Fig.~\ref{fig:var32} indicate 
that the relative amount of states that does reach the typical region essentially increases up to 
 $100\%$ rather 
quickly with system size.
\begin{figure}[htbp]
 \centering
 \includegraphics[width=7cm,height=5cm]{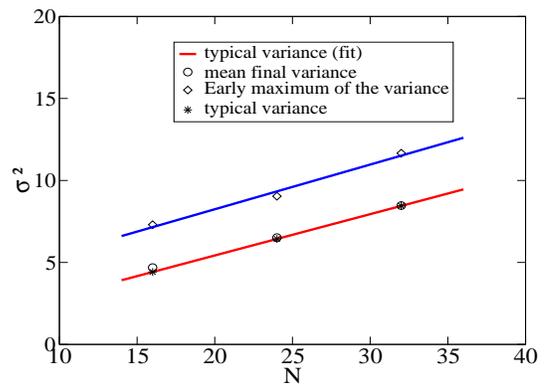}
 \caption{(color online). Scaling of three quantities concerning variances with system size $N$:  ($\circ$), average over all addressed final 
 variances as displayed, e.g., in Figs. \ref{fig:rate32}, \ref{fig:var32}, ($\ast$) typical variances (see text), ($\diamond$) largest early
 maxima as visible, e.g., in   Figs. \ref{fig:rate32}, \ref{fig:var32}. Solid lines are linear fits to the typical variances (red) and the early maxima (blue).
 The coincidence of the mean variances with the typical variances is very good for larger $N$. All quantities appear to scale linearly with system 
 size, the early maxima are above the mean final variances by a constant shift of $\approx 3$} 
 \label{fig:varvsn}
\end{figure}

Next we address the question how the final variances and their maxima (as visible, e.g., in  Fig.~\ref{fig:var32}) scale with the system size.
Since  the predictability of the quantum dynamics by (\ref{rates}) does not fully hold this scaling is crucial for the claim that this 
model shows irreversible, macroscopically deterministic behaviour.
In Fig.~\ref{fig:varvsn} the mean of the final variances averaged over all discussed initial states, the typical variances and 
the largest maxima of the variances are shown for $N=16,24,32$. These maxima occur before the eventual values are reached and are most 
pronounced for the 
respective most ``off-equilibrium'' initial states, i.e., the states with the largest $|\langle\hat{x}(0) \rangle|$, cf. 
Figs.~ \ref{fig:rate32}, \ref{fig:var32}. These early, most pronounced maxima appear to be 
above the final variances by a finite shift of $\approx 3$ for all system sizes as may be inferred from  Fig.~\ref{fig:varvsn}. 
Figure \ref{fig:varvsn} also indicates that, at least for $N=24,32$, the final variances agree very well with the typical variances 
as already mentioned above. It is also clearly seen that all displayed quantities scale linearly with the system size and feature the 
same slope.

The above 
findings may be summarized and interpreted as follows:
Figure \ref{fig:mean32} strongly indicates that the expectation value of the magnetization difference $x$
shows autonomous and Markovian dynamics. Figure \ref{fig:var32} indicates that its variances may go through early maxima but eventually 
tightly cluster around the typical variance. 
Figure \ref{fig:varvsn} indicates that throughout the dynamics the standard deviations  $\sigma$ of the magnetization difference essentially 
scale as $\sigma \propto \sqrt{N}$. Since the maximum   $\langle \hat{x}\rangle$ scales linearily with the system size as $N$ this means that 
the standard deviation will vanish compared to the overall scale of the expectation values with increasing system size. Therefore the 
specifications  for thermodynamic behaviour given
in the introduction (autonomous and Markovian dynamics for the expectation value and negligible
variances on the scale defined by the expectation values) are met for this system.

\section{stochastic dynamics beyond macroscopic determinism}

\begin{figure}[t]
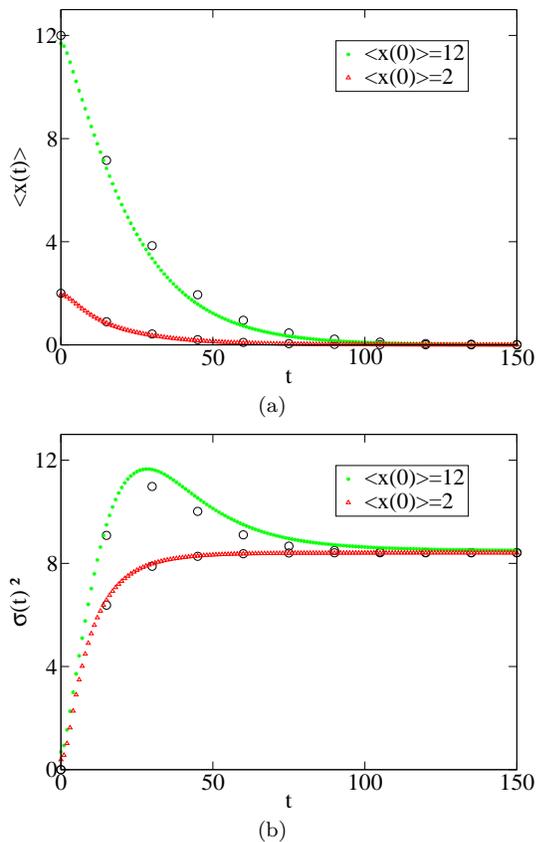

 \centering
\subfigure[]{
   \includegraphics[width=7cm,height=5cm]{fig6a.eps}
\label{fig:ratedisc16}}
\subfigure[]{
\includegraphics[width=7cm,height=5cm]{fig6b.eps}
\label{fig:ratedisc32}}
\caption{(color online). (a) Expectation values $\langle\hat{x}(t) \rangle$ of the magnetization difference 
for $N=32$ for two states, one starting close to equilibrium $\langle \hat{x}(0) \rangle =2$ and the other far from equilibrium  
$\langle \hat{x}(0) \rangle = 12$. Circles ($\circ$) represent the corresponding data as calculated from the Markov chain specified by  
(\ref{trans}). (b) Equivalent data and 
style of presentation as in (a) but for the variances $\sigma^2(t)$.
This is to be compared with Fig.~\ref{fig:N32}. Obviously the agreement of the quantum dynamics with data from Markov chain is better 
than with results from the spin-flip model suggested in Ref.~\cite{Niemeyer2013}.}
 \label{fig:stocuni}
\end{figure}

Considering the correspondence with macroscopic thermodynamic properties the question comes to mind
whether the dynamics of $\langle \hat{x}\rangle$ and $\sigma^2$ can be effectively described by a (discrete) Markov
chain on the magnetization difference subspaces, regardless of the failure of  the model suggested 
in Ref.~\cite{Niemeyer2013} for larger $|X|$. As already pointed out in the above discussion of 
Figs.~\ref{fig:N16}, \ref{fig:N32} this Markov chain should be similar to the model defined by (\ref{rates}) close to 
equilibrium 
but should significantly deviate from the latter for transitions between subspaces with larger $|X|$. In a sense, which is described in 
more detail below, this Markov chain can be expected to be comparable to a Fokker-Planck equation. Moreover,
due to the autonomy of the dynamics of
$\langle \hat{x}(t) \rangle$ this Markov chain must correspond to a Fokker-Planck equation with a drift or force term, the curvature of 
which is negligible on the scale of $\sigma$. The existence of such a Markov chain would imply the validity of what Van Kampen called the 
``assumption of repeated randomness'' \cite{VanKampen1962, Butterfield2007} for this specific system (It may be worth noting here that 
Van Kampen and 
others built their explanation of the second law on this assumption.). Thus, in  order to check whether such a Markov chain 
exists and to find its concrete form we 
compute the finite transition probabilities $w_{XY}(\tau)$ between all subspaces according to :
\begin{equation}
 \label{trans}
 w_{XY}(\tau):= |\hat{P}_{X}  e^{-i\tau \hat{H}}|\omega_Y\rangle|^2. 
 \end{equation}
If such a description applies, 
the dynamics of the probabilities should be given by 
$P_X(t)$ :
\begin{equation}
 \label{iter}
\vec{P}(n\tau)=W(\tau)^{n}\vec{P}(0),
\end{equation} 
where $\vec{P}(t)$ represents the entity of all $\{ P_X(t)\}$ as a vector, and $W(\tau)$ is the transition matrix formed by all 
$\{w_{XY}(\tau)\}$. From  $\vec{P}(t)$ the mean and variance may be computed as 
\begin{equation}
 \label{meanvar}
<x(t)>= \vec{X}\cdot \vec{P}(t),  \quad \sigma^2 = \vec{X^2}\cdot \vec{P}(t)-( \vec{X}\cdot \vec{P}(t))^2,
\end{equation} 
where $ \vec{X}$ is the vector formed by all $\{ X \}$ and $ \vec{X^2}$ the vector formed by all $\{ X^2 \}$. In Fig.~\ref{fig:stocuni} 
we compare the dynamics of mean and variance as resulting from the unitary evolution to the dynamics as resulting from (\ref{trans}, 
\ref{iter}) for $\tau = 15$ (this choice is not imperative, however for $\tau$ on the timescale of the correlation time the agreement
becomes worse).

Obviously there is reasonable agreement even for states starting far from equilibrium. This indicates that a Markov chain
based on $W(\tau)$ may indeed essentially capture the dynamics of the closed quantum system. In order to compare the Markov chain
defined by (\ref{iter}) to the  master equation defined by (\ref{rates}) we first compute a matrix of finite transition probabilities 
$u_{XY}(\tau)$ as resulting from  (\ref{rates}). This is conveniently done numerically. Since we intend to compare with $w_{XY}(\tau=15)$ of 
course 
we compute also $u_{XY}(\tau=15)$. In order to be able to compare $w_{XY}(\tau=15), u_{XY}(\tau=15)$ in a meaningful way we assign to 
both transition 
matrices a ``force'' $f(X)$ and a ``diffusion coefficient'' $D(X)$ in the following way: We compute the change of the mean $<x(t)>$ 
during time $t=15$ given that one started with $<x(0)>=X$, we call that $f(X)$. Furthermore we compute the increase of the variance 
$\sigma^2(t)$ during time $t=15$ given that one started with $<x(0)>=X, \sigma^2(0) \approx 0$, we call that $D(X)$.

\begin{figure}[t]
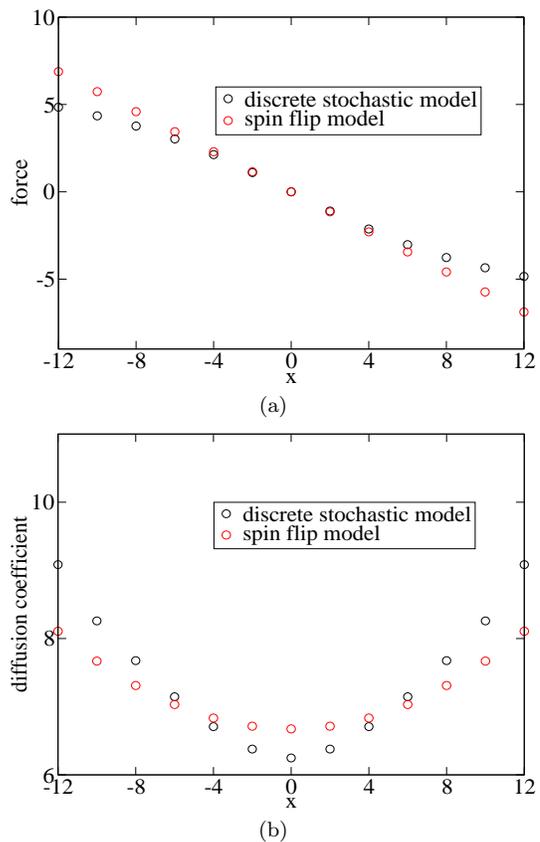

 \centering
\subfigure[]{
   \includegraphics[width=7cm,height=5cm]{fig7a.eps}
\label{fig:force}}
\subfigure[]{
\includegraphics[width=7cm,height=5cm]{fig7b.eps}
\label{fig:diff}}
\caption{(color online). (a) Quantity corresponding to a force term in a Fokker-Planck equation, calculated from the spin-flip model defined 
by (\ref{rates}) (red circles) and  the discrete stochastic model (\ref{trans}) (black circles). While the force for the spin flip model
is almost
strictly linear, the force for the discrete stochastic model deviates from that in the off-equilibrium regime. However, the curvature is low
on the scale of $\sigma$ ($\approx 3$). 
 (b) Equivalent data and display style as in (a) but for a quantity corresponding to the diffusion term in a 
 Fokker-Planck equation. The diffusion coefficient for both models  deviate 
slightly from each other.}
\label{fig:forcdif}
\end{figure}

The results for the forces and the diffusion coefficients are displayed in  Fig.~\ref{fig:forcdif} 
Obviously there is a good agreement for the force and reasonable agreement for the diffusion coefficient as calculated from the Markov chain
(\ref{trans}) and the 
spin-flip model (\ref{rates}) close to equilibrium ($X=0$). However there are also significant differences in the off-equilibrium regime.
Based on the numerics at hand we did not find any tendency of 
these differences to vanish in the limit of larger systems.

\section{summary, conclusion and outlook}

In this paper we have demonstrated that a finite quantum system may show thermodynamic behaviour in the sense of macroscopically
deterministic, autonomous, Markovian
relaxation of an observable for a very large class of  pure initial states. While this behavior gets more pronounced under upscaling it
is already visible for a system comprising $32$ spins. Although being in a well-defined sense Markovian, the above relaxation dynamics is not 
in full accord with the 
naive model presented in Ref. \cite{Niemeyer2013}. However, this finding nonetheless supports the view  of irreversible, stochastic  dynamics of 
selected observables emerging
directly from quantum mechanics. Of course this emergence of thermodynamical 
relaxation 
directly leads to the quest for a sensible definition of entropy in this context. Since (regardless of the failure of the Fokker-Planck based 
model in  Ref. \cite{Niemeyer2013} the dynamics is found to be in accord with a specific Markov chain) such a notion could be provided by 
the concept of stochastic thermodynamics as described, e.g., in Refs \cite {Sekimoto1998, Lebowitz1999, Seifert2005, Esposito2010}. 
(For a comprehensive 
introduction see also \cite{Seifert2012}.) This concept has already been applied to Markovian, \cite{Esposito2006} and 
non-Markovian \cite{Kawamoto2011} open quantum systems. The model presented in the paper at hand could provide an access to a systematic application 
of stochastic thermodynamics to closed quantum systems.

This work is supported in part by NCF, The Netherlands (HDR).
The authors gratefully acknowledge the computing time granted by the JARA-HPC Vergabegremium and
provided on the JARA-HPC Partition part of the supercomputer JUQUEEN at Forschungszentrum J\"ulich.


%

\end{document}